\documentclass[12pt]{article}
\usepackage{amssymb,amsmath}
\begin{document}


\sloppy
\title
{\Large  Problems in field theoretical approach to gravitation }

\author
 {
       A.I.Nikishov
          \thanks
             {E-mail: nikishov@lpi.ru}
  \\
               {\small \phantom{uuu}}
  \\
           {\it {\small} I.E.Tamm Department of Theoretical Physics,}
  \\
               {\it {\small} P.N.Lebedev Physical Institute, Moscow, Russia}
  \\
 }
%
\maketitle
\begin{abstract}
 We consider gravitational self interaction in the lowest approximation
 and assume that graviton interacts with gravitational energy-momentum tensor
 in the same way as it interacts with particles. We show that, using
 gravitational vertex with a preferred gravitational  energy-momentum tensor,
 it is possible to obtain a metric necessary for explaining perihelion 
 precession. The preferred gravitational energy-momentum tensor gives positive
 gravitational energy density of Newtonian center. We show also that, employing
 "improvement" technique, any gravitational energy-momentum tensor can be made
 suitable for using in gravitational wave equation for obtaining metric which
 explains perihelion precession. Yet the "improvement" leads to negative
 gravitational energy density of the Newtonian center.
\end{abstract}

\section{Introduction}

The field theoretical approach to gravitation gives an additional insight
into gravity both within general relativity and beyond. For example,
this approach can help to choose the preferred coordinate system, which
 enable as to
see directly from $g_{\mu\nu}$ how the gravitational field affects the 
Cartesian latticework of rigid rods, existed before switching on gravitational
field. The field theoretical approach suggests that in a weak gravitational
field of a Newtonian center the influence on rods is isotropic [Thirring (1961)].
The same conclusion follows from the dimensional analysis in 
[Dehnen, H\"onl, Westpfahl 
 (1960)]. On the other hand, the standard Schwarzschild coordinate system 
 suggests that only radial lengths are affected by spherically symmetric 
 gravitational field, cf. for example, [Cohn (1969)], see also \S 22 in
 [Einstein (1916)]. Yet, in the framework of general relativity we 
 cannot get the answer to this question without performing first the
 complicated procedure of identification coordinates $x$ in $g_{\mu\nu}(x)$ 
with points in the laboratory, cf. \S 22 in [Misner, Thorne, Wheeler  (1973)].

If the influence of gravitational field is indeed anisotropic, then the
atom in gravitational field should lose its spherical symmetry and this must
affect its spectrum, cf. [Thirring (1961)]. 
We note also that the ability to define the preferred system can also help us
 to resolve the problem of nonuniqueness of gravitational energy-momentum 
 tensor [Nikishov (2003)].

 It is commonly accepted that using field theoretical methods in building up
 gravity theory one almost inevitably end up with general relativity,
 see, for example, [Wald (1986)]. We show that at least in the lowest nonlinear
 approximation this is not the case. One appealing way to get a version 
 of gravity theory is to use the $S-$matrix formalism with propagators 
 and vertices without any recourse to gravitational wave equation. This
 way we are free from the assumption that there are purely gravitational
 Lagrangian and matter Lagrangian; they may be inextricable.

 We consider the 3-graviton vertex build with the help of a preferred
 gravitational energy-momentum tensor which yields positive gravitational 
 energy density of the Newtonian center. Using this vertex it is possible 
 to obtain metric necessary for explanation of perihelion precession.

 It seems natural to assume that gravitational field
 should have positive energy density. In this case the existence of
 black holes should be impossible. Some arguments in favor of this are
 given in [Baryshev (1999)]. It is likely that in the future the
  gravitational theory
 will be able to define uniquely the gravitational energy-momentum 
 tensor.

 In my previous paper [Nikishov (1999)] I have considered several gravitational  
 energy-momentum tensors. Only one of them gives positive gravitational
  energy density of the Newtonian center. In this paper I use this tensor
 in 3-graviton vertex.

 In Section 2 a notation is introduced for the building blocks of gravitational 
 energy-momentum tensors. It facilitates the comparison of different tensors
 and different theories. In Section 3 we  obtain the necessary 
 $g_{\mu\nu}$ from our vertex. In Section 4 we show how to get different 
 versions of  gravitational wave equations by using "improvement" technique
 for gravitational energy-momentum tensors. In Section 5 we compare the
 Feynman's gravitational Lagrangian [Feynman Moringo, Wagner (1995)] with that
  of general relativity. 

  \section{Notation and some preliminary relations}

  We define
    $$
  g_{\mu\nu}=\eta_{\mu\nu}+h_{\mu\nu},\quad \bar h_{\mu\nu}=h_{\mu\nu}-
  \frac12\eta_{\mu\nu}h,\quad h=h_{\alpha}{}^{\alpha},
  \quad \eta_{\mu\nu}={\rm diag}(-1,1,1,1).                            \eqno(1)
  $$
  If not said otherwise Latin and Greek indices run from 0 to 3.
  As we are going to compare different gravitational energy-momentum tensors,
  it is helpful to introduce special notation for the building blocks of these 
  tensors. We denote
  $$
\stackrel{1}\tau{}^{jk}=\eta^{jk}h_{\alpha\beta,\gamma}h^{\gamma\beta,\alpha};
\quad  \stackrel{2}\tau{}^{jk}=
\eta^{jk}h_{\alpha\beta,\gamma}h^{\alpha\beta,\gamma};
\quad\stackrel{3}\tau{}^{jk}=\eta^{jk}h_{,\rho}h^{\rho\sigma}{},_{\sigma}
\quad\stackrel{4}\tau{}^{jk}=\eta^{jk}h_{,\rho}h^{,\rho};
  $$
 $$
 \stackrel{5}\tau{}^{jk}=h_{\alpha\beta}{}^{,j}h^{\alpha\beta,k};
 \quad \stackrel{6}\tau{}^{jk}=h^{j\alpha,\beta}h^{k}{}_{\beta,\alpha};\quad
 \stackrel{7}\tau{}^{jk}=h^{j\alpha,\beta}h^{k}{}_{\alpha,\beta};\quad
 \stackrel{8}\tau{}^{jk}=\frac12(h^{j\alpha,k}+h^{k\alpha,j})h_{,\alpha};
 $$
   $$
\stackrel{9}\tau{}^{jk}=h^{jk,\sigma}h_{,\sigma};\quad\stackrel{10}\tau{}^{jk}=
\frac12(h^{j\sigma}{}_{,\sigma}h^{,k}+h^{k\sigma}{}_{,\sigma}h^{,j});\quad
\stackrel{11}\tau{}^{jk}=h^{,j}h^{,k};\quad\stackrel{12}\tau{}^{jk}=
h^{jk,\alpha}h_{\alpha\sigma}{}^{,\sigma};
   $$
   $$
\stackrel{13}\tau{}^{jk}=\frac12(h^{j\alpha,\beta}h_{\alpha\beta}{}^{,k}
+h^{k\alpha,\beta}h_{\alpha\beta}{}^{,j});\quad\stackrel{14}\tau{}^{jk}=
h^{j\sigma}{}_{,\sigma}h^{k\alpha}{}_{,\alpha};
\quad \stackrel{15}\tau{}^{jk}=\frac12(h^{j\alpha,k}+h^{k\alpha,j})
h_{\alpha\sigma}{}^{,\sigma};
   $$
 $$
\stackrel{16}\tau{}^{jk}=\eta^{jk}h_{\alpha\beta}{}^{,\beta}h^{\alpha\sigma}
{_{,\sigma}};\quad h_{,i}=\frac{\partial h}
{\partial x^i}.                                                     \eqno(2)
 $$
Similarly for terms with second derivative:
$$
\stackrel{a}\tau{}^{jk}=\eta^{jk}h_{,\sigma}{}^{\sigma}h;\quad
\stackrel{b}\tau{}^{jk}=\eta^{jk}h_{\alpha\beta}{}^{,\alpha\beta}h;\quad
\stackrel{c}\tau{}^{jk}=\eta^{jk}h^{,\alpha\beta}h_{\alpha\beta};\quad
\stackrel{d}\tau{}^{jk}=
\eta^{jk}h^{\alpha\beta,\sigma}{}_{\sigma}h_{\alpha\beta};
$$
$$
\stackrel{e}\tau{}^{jk}=\eta^{jk}h^{\alpha\sigma}{}_{,\sigma}{}^{\beta}
h_{\alpha\beta};\quad
\stackrel{f}\tau{}^{jk}=h^{,jk}h;\quad \stackrel{g}\tau{}^{jk}=h^{jk,\sigma}
{}_{\sigma}h;\quad \stackrel{h}\tau{}^{jk}=\frac12(h^{j\sigma,k}{}_{\sigma}+
h^{k\sigma,j}{}_{\sigma})h;
$$
$$ 
  \quad\stackrel{i}\tau{}^{jk}=h^{jk,\alpha\beta}h_{\alpha\beta};\quad
  \stackrel{j}\tau{}^{jk}=\frac12(h^{j\alpha,k\beta}+
 h^{k\alpha,j\beta})h_{\alpha\beta};\quad\stackrel{k}\tau{}^{jk}=
 h^{\alpha\beta,jk}h_{\alpha\beta}; \quad\stackrel{l}\tau{}^{jk}=
 h^{jk}h_{,\sigma}{}^{\sigma};
  $$
  $$
  \quad\stackrel{m}\tau{}^{jk}=
 h^{jk}h_{\alpha\beta}{}^{,\alpha\beta};\quad \stackrel{n}\tau{}^{jk}=
 \frac12(h^{,j\alpha}h_{\alpha}{}^{k}+h^{,k\alpha}h_{\alpha}{}^{j});\quad
 \stackrel{o}\tau{}^{jk}=
 \frac12(h^{j\sigma,\alpha}{}_{\sigma}h_{\alpha}{}^{k}+
 h^{k\sigma,\alpha}{}_{\sigma}h_{\alpha}{}^{j});
  $$
  $$
 \stackrel{p}\tau{}^{jk}=
 \frac12(h^{j\alpha,\sigma}{}_{\sigma}h_{\alpha}{}^{k}+
 h^{k\alpha,\sigma}{}_{\sigma}h_{\alpha}{}^{j});\quad\stackrel{q}\tau{}^{jk}=
 \frac12(h^{\alpha\sigma,j}{}_{\sigma}h_{\alpha}{}^{k}+
 h^{\alpha\sigma,k}{}_{\sigma}h_{\alpha}{}^{j}).                    \eqno(3)
  $$
  Tensors $\stackrel{1}{\cal T}{}^{jk},\cdots \stackrel{16}{\cal T}{}^{jk}$,
   and $\stackrel{a}{\cal T}{}^{jk},\cdots\stackrel{q}{\cal T}{}^{jk}$  are 
   obtained from (1) and (2) by substitutions
   $h_{\mu\nu}\to\bar h_{\mu\nu}$.
  There are relations
  $$
  \stackrel{a}{\cal T}{}^{jk}=\stackrel{a}\tau{}^{jk};\quad
  \stackrel{b}{\cal T}{}^{jk}=
  \frac12\stackrel{a}\tau{}^{jk}-\stackrel{b}\tau{}^{jk};
  \quad \stackrel{c}{\cal T}{}^{jk}=
  \frac12\stackrel{a}\tau{}^{jk}-\stackrel{c}\tau{}^{jk};\quad
  \stackrel{d}{\cal T}{}^{jk}=\stackrel{d}\tau{}^{jk};
  $$
   $$
   \stackrel{e}{\cal T}{}^{jk}=
  \frac14\stackrel{a}\tau{}^{jk}-\frac12\stackrel{b}\tau{}^{jk}
  -\frac12\stackrel{c}\tau{}^{jk}+\stackrel{e}\tau{}^{jk};\quad
  \stackrel{f}{\cal T}{}^{jk}=\stackrel{f}\tau{}^{jk};\quad
  \stackrel{g}{\cal T}{}^{jk}=\frac12\stackrel{a}\tau{}^{jk}-
  \stackrel{g}\tau{}^{jk};\quad
  \stackrel{h}{\cal T}{}^{jk}=\frac12\stackrel{f}\tau{}^{jk}
  -\stackrel{h}\tau{}^{jk};
   $$
   $$
   \stackrel{i}{\cal T}{}^{jk}=\frac14\stackrel{a}\tau{}^{jk}-
  \frac12\stackrel{c}\tau{}^{jk}-\frac12\stackrel{g}\tau{}^{jk}
  +\stackrel{i}\tau{}^{jk};\quad
   \stackrel{j}{\cal T}{}^{jk}=\frac14\stackrel{f}\tau{}^{jk}
   -\frac12\stackrel{h}\tau{}^{jk}+\stackrel{j}\tau{}^{jk}
 -\frac12\stackrel{n}\tau{}^{jk};
  $$
 $$
 \stackrel{k}{\cal T}{}^{jk}=\stackrel{k}\tau{}^{jk};\quad
 \stackrel{l}{\cal T}{}^{jk}=\frac12\stackrel{a}\tau{}^{jk}-
 \stackrel{l}\tau{}^{jk};\quad
  \stackrel{m}{\cal T}{}^{jk}=\frac14\stackrel{a}\tau{}^{jk}-
  \frac12\stackrel{b}\tau{}^{jk}-\frac12\stackrel{l}\tau{}^{jk}
 +\stackrel{m}\tau{}^{jk};
  $$
 $$
 \stackrel{n}{\cal T}{}^{jk}=\frac12\stackrel{f}\tau{}^{jk}
 -\stackrel{n}\tau{}^{jk};\quad \stackrel{o}{\cal T}{}^{jk}=
  \frac14\stackrel{f}\tau{}^{jk}-\frac12\stackrel{h}\tau{}^{jk}
  -\frac12\stackrel{n}\tau{}^{jk}
  +\stackrel{o}\tau{}^{jk};
 $$
   $$
   \stackrel{p}{\cal T}{}^{jk}=\frac14\stackrel{a}\tau{}^{jk}
  -\frac12\stackrel{g}\tau{}^{jk}-\frac12\stackrel{l}\tau{}^{jk}
  +\stackrel{p}\tau{}^{jk};\quad \stackrel{q}{\cal T}{}^{jk}=
  \frac14\stackrel{f}\tau{}^{jk}
  -\frac12\stackrel{h}\tau{}^{jk}-\frac12\stackrel{n}\tau{}^{jk}
  +\stackrel{q}\tau{}^{jk},                                         \eqno(4)
  $$
  and these
  $$
  \stackrel{1}{\cal T}{}^{jk}=
 \stackrel{1}\tau{}^{jk}-
  \stackrel{3}\tau{}^{jk}+
  \frac14\stackrel{4}\tau{}^{jk};\quad \stackrel{2}{\cal T}{}^{jk}=
 \stackrel{2}\tau{}^{jk};\quad \stackrel{3}{\cal T}{}^{jk}=
 -\stackrel{3}\tau{}^{jk}+
  \frac12\stackrel{4}\tau{}^{jk};\quad \stackrel{4}{\cal T}{}^{jk}=
 \stackrel{4}\tau{}^{jk};
  $$
$$
\quad \stackrel{5}{\cal T}{}^{jk}=
 \stackrel{5}\tau{}^{jk}; \quad\stackrel{6}{\cal T}{}^{jk}=
 \stackrel{6}\tau{}^{jk}-
  \stackrel{8}\tau{}^{jk}+
  \frac14\stackrel{11}\tau{}^{jk};\quad\stackrel{7}{\cal T}{}^{jk}=
 \frac14\stackrel{4}\tau{}^{jk}+\stackrel{7}\tau{}^{jk}-
  \stackrel{9}\tau{}^{jk};\quad\stackrel{8}{\cal T}{}^{jk}=
 -\stackrel{8}\tau{}^{jk}+
  \frac12\stackrel{11}\tau{}^{jk};
 $$
   $$
   \stackrel{9}{\cal T}{}^{jk}=\frac12\stackrel{4}\tau{}^{jk}
 -\stackrel{9}\tau{}^{jk}
 ;\quad \stackrel{10}{\cal T}{}^{jk}=
 -\stackrel{10}\tau{}^{jk}+
  \frac12\stackrel{11}\tau{}^{jk};\quad   \stackrel{11}{\cal T}{}^{jk}=
 \stackrel{11}\tau{}^{jk};
 $$
$$
\stackrel{12}{\cal T}{}^{jk}=-
  \frac12\stackrel{3}\tau{}^{jk}+
  \frac14\stackrel{4}\tau{}^{jk}-\frac12\stackrel{9}\tau{}^{jk}
  +\stackrel{12}\tau{}^{jk};\quad\stackrel{13}{\cal T}{}^{jk}=-
  -\frac12\stackrel{8}\tau{}^{jk}-\frac12\stackrel{10}\tau{}^{jk}+
  \frac14\stackrel{11}\tau{}^{jk}+\stackrel{13}\tau{}^{jk};
  $$
    $$
    \stackrel{14}{\cal T}{}^{jk}=
  -\stackrel{10}\tau{}^{jk}+\frac14\stackrel{11}\tau{}^{jk}+
  \stackrel{14}\tau{}^{jk};\quad
  \stackrel{15}{\cal T}{}^{jk}=
  -\frac12\stackrel{8}\tau{}^{jk}-\frac12\stackrel{10}\tau{}^{jk}+
  \frac14\stackrel{11}\tau{}^{jk}+\stackrel{15}\tau{}^{jk};
  $$
$$
  \stackrel{16}{\cal T}{}^{jk}=
  -\stackrel{3}\tau{}^{jk}+\frac14\stackrel{4}\tau{}^{jk}
  +\stackrel{16}\tau{}^{jk}.                                        \eqno(5)
$$
The reversed relations are obtained from (4) and (5) by substitutions
${\cal T}\leftrightarrow\tau$.

Next we introduce the Newtonian potential $\phi$. For one Newtonian center
$\phi=-\frac{GM}{r}$, for several centers
$$
\phi=-G\sum_a\frac{m_a}{\vert\vec r-\vec r_a\vert}.                \eqno (6)
$$
For these centers the field theoretical approach gives in linear
 approximation [Thirring (1961)]
$$
\bar h_{\mu\nu}^{(1)}=-4\phi\delta_{\mu o}\delta_{\nu 0},\quad
 h_{\mu\nu}^{(1)}=-2\phi\delta_{\mu \nu},\quad h^{(1)}
 \equiv h^{(1)}_{\sigma}{}^{\sigma}=-4\phi=-\bar h^{(1)},
 \quad \eta_{\mu\nu}
={\rm diag}(-1,1,1,1).                                            \eqno(7)
$$
The Schwarzschild metric in harmonic and isotropic (but not in standard)
frames gives (7) in linear approximation. The same expressions is
 obtained by a heuristic method [Dehnen, H\"onl, and Westpfahl (1960)].
Using (7), for the Newtonian centers we find for ${\cal T}^{jk}$
$$
 \stackrel{1}{\cal T}{}^{jk}=\stackrel{3}{\cal T}{}^{jk}=
 \stackrel{6}{\cal T}{}^{jk}=\stackrel{8}{\cal T}{}^{jk}=
 \stackrel{10}{\cal T}{}^{jk}=\stackrel{12}{\cal T}{}^{jk}=
 \stackrel{13}{\cal T}{}^{jk}=\stackrel{14}{\cal T}{}^{jk}=
 \stackrel{15}{\cal T}{}^{jk}=\stackrel{16}{\cal T}{}^{jk}=0;
 $$
   $$
   \stackrel{2}{\cal T}{}^{jk}=\stackrel{4}{\cal T}{}^{jk}=16\eta^{jk}
   (\nabla\phi)^2;\quad \stackrel{5}{\cal T}{}^{jk}=
   \stackrel{11}{\cal T}{}^{jk}=16\phi_{,j}\phi_{,k};\quad
  \stackrel{7}{\cal T}{}^{jk}=\stackrel{9}{\cal T}{}^{jk}=
  -16(\nabla\phi)^2\delta_{j0}\delta_{k;0};
   $$
 $$
   \stackrel{a}{\cal T}{}^{jk}=\stackrel{d}{\cal T}{}^{jk}=16\eta^{jk}
   \phi\phi_{,ii};\quad  \stackrel{g}{\cal T}{}^{jk}
   =\stackrel{l}{\cal T}{}^{jk}=\stackrel{p}{\cal T}{}^{jk}=
   -16\phi\phi_{,ii}\delta_{j0}\delta_{k;0};\quad
   \stackrel{f}{\cal T}{}^{jk}=\stackrel{k}{\cal T}{}^{jk}=
   16\phi\phi_{,jk};
  $$
  $$
  \stackrel{b}{\cal T}{}^{jk}=\stackrel{c}{\cal T}{}^{jk}=
  \stackrel{e}{\cal T}{}^{jk}=\stackrel{h}{\cal T}{}^{jk}=
  \stackrel{i}{\cal T}{}^{jk}=\stackrel{j}{\cal T}{}^{jk}=
  \stackrel{m}{\cal T}{}^{jk}=\stackrel{n}{\cal T}{}^{jk}=
  \stackrel{o}{\cal T}{}^{jk}=\stackrel{q}{\cal T}{}^{jk}=0,          \eqno(8)
  $$
  and for $\tau^{jk}$
 $$
 \stackrel{1}\tau{}^{jk}=\frac14\stackrel{2}\tau{}^{jk}=
 \frac12\stackrel{3}\tau{}^{jk}=\frac14\stackrel{4}\tau{}^{jk}=
 \stackrel{7}\tau{}^{jk}=\stackrel{16}\tau{}^{jk}=
 4\eta^{jk}(\nabla\phi)^2;\quad \stackrel{5}\tau{}^{jk}=
 4\stackrel{6}\tau{}^{jk}=
 $$
 $$
 2\stackrel{8}\tau{}^{jk}=
 2\stackrel{10}\tau{}^{jk}=\stackrel{11}\tau{}^{jk}=
 4\stackrel{13}\tau{}^{jk}=4\stackrel{14}\tau{}^{jk}=
 4\stackrel{15}\tau{}^{jk}=16\phi_{,j}\phi_{,k};\quad\stackrel{9}\tau{}^{jk}=
 2\stackrel{12}\tau{}^{jk}=8\delta_{jk}(\nabla\phi)^2;
 $$
 $$
 \frac12\stackrel{a}\tau{}^{jk}=\stackrel{b}\tau{}^{jk}=
 \stackrel{c}\tau{}^{jk}=\frac12\stackrel{d}\tau{}^{jk}=
 2\stackrel{e}\tau{}^{jk}=2\stackrel{p}\tau{}^{jk}=
 8\eta^{jk}\phi\phi_{,ii};\quad
 \stackrel{g}\tau{}^{jk}=2\stackrel{i}\tau{}^{jk}=
 \stackrel{l}\tau{}^{jk}=
 $$
   $$
   2\stackrel{m}\tau{}^{jk}=8\delta_{jk}\phi\phi_{,ii};\quad
 \frac12\stackrel{f}\tau{}^{jk}=\stackrel{h}\tau{}^{jk}=
 2\stackrel{j}\tau{}^{jk}=\frac12\stackrel{k}\tau{}^{jk}=
 \stackrel{n}\tau{}^{jk}=2\stackrel{o}\tau{}^{jk}=2\stackrel{q}\tau{}^{jk}
 =8\phi\phi_{,jk}.                                                   \eqno(9)
 $$
 $j,k=0,1,2,3.$
 It is assumed in (8) and (9) that $\eta_{jk}={\rm diag}(-1,1,1,1)$. Otherwise
 one mast replace $\eta_{jk}$ by $-\eta_{jk}$. 

 \section{Gravitational energy-momentum tensor}

 In my previous paper [Nikishov (1999)] I have considered several 
 gravitational 
 energy-momentum tensors in the lowest (e.i. $h^2$) approximation. One
 of them (corresponding to general relativity) is obtainable from the
  Lagrangian
 $$
 \stackrel{Thir}L=\frac{1}{32\pi G}[-\frac12\bar h_{\alpha\beta,\gamma}
 \bar h^{\alpha\beta,\gamma}+ \frac14\bar h_{,\alpha}\bar h^{,\alpha}+
 \bar h_{\alpha\beta,\lambda}\bar h^{\lambda\beta,\alpha}],
 $$
 the other one from
 $$
 \stackrel{MTW}L=\frac{1}{32\pi G}[-\frac12\bar h_{\alpha\beta,\gamma}
 \bar h^{\alpha\beta,\gamma}+ \frac14\bar h_{,\alpha}\bar h^{,\alpha}+
 \bar h_{\alpha\beta}{}^{,\alpha}\bar h^{\beta\lambda}{}_{,\lambda}].\eqno(10)
 $$
 (The Lagrangian (10) is eq. (7.8b) in [Misner, Thorne, Wheeler (1973)];
 earlier it was used by Feynman, see eqs. (3.6.1) and (3.6.5) in 
 [Feynman, Moringo, Wagner (1995)].) 
 These Lagrangians are related as follows
 $$
 \stackrel{Thir}L=\stackrel{MTW}L+\frac1{32\pi G}[(
 \bar h_{\alpha\beta}\bar h^{\lambda\beta,\alpha})_{,\lambda}-
 (\bar h_{\alpha\beta}\bar h^{\lambda\beta}{}_{,\lambda})^{,\alpha}].
 $$
  Symmetrizing the canonical energy-momentum tensor,
  obtained from (10) by
 Belinfante method and adding the interaction tensor\\ $\stackrel{int}T{}^{jk}=
 \stackrel{M}T{}^{jn}h_n{}^k$, we obtain [Nikishov (1999)]
$$
t^{jk}=\frac1{32\pi G}[-\frac12\stackrel{2}{\cal T}{}^{jk} +
\frac14\stackrel{4}{\cal T}{}^{jk}+\stackrel{5}{\cal T}{}^{jk}-
\frac12\stackrel{11}{\cal T}{}^{jk}+2\stackrel{12}{\cal T}{}^{jk}-
2\stackrel{13}{\cal T}{}^{jk}-
2\stackrel{14}{\cal T}{}^{jk}+
$$
$$
2\stackrel{15}{\cal T}{}^{jk}-\stackrel{16}{\cal T}{}^{jk}+
2\stackrel{j}{\cal T}{}^{jk}-2\stackrel{e}{\cal T}{}^{jk}+
2\stackrel{m}{\cal T}{}^{jk}-2\stackrel{o}{\cal T}{}^{jk}]+
\frac12(\stackrel{M}T{}^{jn}h_n{}^k+\stackrel{M}T{}^{kn}h_n{}^j).    \eqno(11)
$$
Here $\stackrel{M}T{}^{jk}$ is energy-momentum tensor of particles
   $$
 \stackrel{M}{T^{jk}}=
\sum_am_au^{j}u^{k}\frac{ds}{dt}\delta(\vec x-\vec x_a(t)),\quad
u^{\mu}=dx^{\mu}/ds.                                             \eqno (12)
$$
The remarkable feature of tensor (11), called in [Nikishov (1999)]
 MTW-tensor, is that
it gives positive energy density of gravitational field of the Newtonian center.
Moreover, the expression for gravitational energy density has the same form
 as electromagnetic energy density of a Coulomb center:
 $$
 t^{00}=2\stackrel{M}{T^{00}}\phi+\frac1{8\pi G}(\nabla\phi)^2. \eqno(13)
 $$ 
 Together with (12) we find for slowly moving particles 
 $$
 \stackrel{M}{T^{00}}+t^{00}=\sum_am_a\delta(\vec x-\vec x_a(t))+
\stackrel{M}{T^{00}}\phi+\frac1{8\pi G}(\nabla\phi)^2.                 \eqno(14)
$$
Here we have used eq.(29) below and eq.(7) according to which $h_{00}=-2\phi$.
Now we consider only two particles $m_a$ and $m_b$ and drop the self 
interaction terms in $\stackrel{M}{T^{00}}\phi$:
$$
-[m_a\delta(\vec r-\vec r_a)+m_b\delta(\vec r-\vec r_b)][
\frac{Gm_a}{|\vec r-\vec r_a|}+\frac{Gm_b}{|\vec r-\vec r_b|}]
$$
$$
\to-\frac{Gm_am_m}{|\vec r_a-\vec r_b|}\delta(\vec r-\vec r_a)-
\frac{Gm_am_m}{|\vec r_a-\vec r_b|}\delta(\vec r-\vec r_b).         \eqno(15)
$$
Then, the probing particle $m_a$, slowly moving in the gravitational field of
 $m_b$, has the energy 
  $m_a\left(1-\frac{Gm_b}{|\vec r_a-\vec r_b|}\right)$,
  see terms proportional to $\delta(\vec r-\vec r_a)$ in (14) and (15).
   This is what is needed
  for explaining the red shift according to \S4 in Ch 2 in [Schwinger (1970)].
Indeed, let $w_{\infty}$ be the frequency of an atom outside the gravitational
field. Then $w_{\infty}(1+\phi(r_a))$ is the frequency in the field in terms of
(coordinate) time $t$. In terms of observable time $t^{obs}=\sqrt{|g_{00}|}t$ at
 point $r$ the observable frequency
 is $w_{\infty}(1+\phi(r_a))(1-\phi(r))$. In particular, near the point 
 of emission $r=r_a$ the observable frequency is $w_{\infty}$, i.e. the same as
 without gravitational field. This is in agreement with Thirring (1961),
 who considered electron-proton system in gravitational field, using
 observable length and $t^{obs}$. 
                                  
We see that tensor (11) may be the  real gravitational
energy-momentum tensor and all other tensors are only effective tensors,
 leading to correct expressions for special cases such as total energy of 
 a system, or energy density of gravitational wave.   
   
 In view of this properties we try to use MTW-tensor in three graviton 
 vertex. The immediate aim is to obtain $g_{00}$ for the Newtonian 
 center  in the form
  $$
 g_{00}=-(1+2\phi+2\phi^2)                                         \eqno(16)
 $$
 necessary for explaining  the perihelion precession of Mercury, if the linear 
 approximation  for $g_{\mu\nu}=\eta_{\mu\nu}+h_{\mu\nu}$ is given by (7).
  So we use the gravitation interaction Lagrangian in the form
 $$
 \stackrel{int}L=\frac12h_{\mu\nu}t^{\mu\nu}                      \eqno(17)
 $$ 
 with $t^{\mu\nu}$ given in (11).

 The corresponding general relativity 
 Lagrangian in terms of $\tau^{\mu\nu}$ is (see [Iwasaki (1971)])
 $$
 \frac1{64\pi G}h_{\mu\nu}\sum_{s=1}^{13}a_s\stackrel{s}\tau{}^{\mu\nu}.
 \quad                                                              \eqno(18)
 $$
 Here $a_s$ are given in Table 1
 \begin{center}
Table 1
\end{center}
$$
\begin{array}{|c|c|c|c|c|c|c|c|c|c|c|c|c|c|}
\hline
s&1&2&3&4&5&6&7&8&9&10&11&12&13\\
\hline
a_s&1&-\frac12&-1&\frac12&1&-2&2&2&-2&2&-1&2&-4 \\
\hline
\end{array}
$$
(In [Iwasaki (1971)] $a_{11}=-2$; this must be a mistake.)

If we rewrite (16) in the form
  $$
 g_{00}=-[1+2(\phi+c\phi^2)],                           \eqno(19)
 $$
 then we can find from Iwasaki paper the 
contribution $c_s=a_sR_s$ to the sum $c=\sum_1^{13}c_s$ for each
 $s=1,2,\cdots,13$. Hence we can find $R_s$ and use them in different 
 versions of theory. So we get $R_s$ given in Table 2.
 \begin{center}
Table 2
\end{center}
$$
\begin{array}{|c|c|c|c|c|c|c|c|c|c|c|c|c|c|c|c|c|}
\hline
s&1&2&3&4&5&6&7&8&9&10&11&12&13&14&15&16\\
\hline
R_s&-\frac12&-2&-1&-2&-1&-\frac14&-\frac12&-\frac12&-2&-\frac12&-1&-1&-\frac14
&-\frac14&-\frac14&-\frac12 \\
\hline     e_s&1&-\frac12&-1&\frac12&1&2&0&1&1&3&-2&-2&-2
&-2&0&0 \\
\hline
\end{array}
$$
The table is enlarged because in general $s=14, 15, 16$ also occur. 
We note that $R_s$ is determined by Newtonian value of 
$\stackrel{s}\tau{}^{\mu\nu}$. For example, from $\stackrel{1}\tau{}^{\mu\nu}=
\stackrel{7}\tau{}^{\mu\nu}=\stackrel{16}\tau{}^{\mu\nu}$ in (9) it follows
$R_1=R_7=R_{16}$ and so on.

To use Table 2 for the case with three graviton vertex (17),
 we rewrite $t^{jk}$ in (11)  in terms of $\tau$. With the 
help of (4) and (5) we find
$$
t^{jk}=\frac1{32\pi G}[-\frac12\stackrel{2}{\tau}{}^{jk} +
\frac12\stackrel{4}{\tau}{}^{jk}+\stackrel{5}{\tau}{}^{jk}-
\stackrel{9}{\tau}{}^{jk}+2\stackrel{10}{\tau}{}^{jk}-
\stackrel{11}{\tau}{}^{jk}+2\stackrel{12}{\tau}{}^{jk}-
2\stackrel{13}{\tau}{}^{jk}-
2\stackrel{14}{\tau}{}^{jk}+
$$
$$
2\stackrel{15}{\tau}{}^{jk}-\stackrel{16}{\tau}{}^{jk}+
\stackrel{c}{\tau}{}^{jk}-
2\stackrel{e}{\tau}{}^{jk}+2\stackrel{j}{\tau}{}^{jk}-
\stackrel{l}{\tau}{}^{jk}+
2\stackrel{m}{\tau}{}^{jk}-2\stackrel{o}{\tau}{}^{jk}]+
\frac12(\stackrel{M}T{}^{jn}h_n{}^k+\stackrel{M}T{}^{kn}h_n{}^j).    \eqno(20)
$$

  Next, we convert the terms with second derivatives
  into terms with first derivatives.
  For the term $\stackrel{a}{\tau}{}^{\mu\nu}$ we have
  $$
  h_{\mu\nu}\stackrel{a}{\tau}{}^{\mu\nu}\equiv h_{\mu\nu}\eta^{\mu\nu}
  h_{,\sigma}{}^{,\sigma}h=(h^{,\sigma}hh)_{,\sigma}-
  h^{,\sigma}2hh_{,\sigma}.                                        \eqno(21)
  $$
   Dropping divergence, we write $h_{\mu\nu}\stackrel{a}{\tau}{}^{\mu\nu}
   \to-2h_{\mu\nu}\stackrel{4}\tau{}^{\mu\nu}$, or symbolically
 $\stackrel{a}\tau{}^{\mu\nu}\to-2\stackrel{4}\tau{}^{\mu\nu}$
 and similarly for other terms.
 Thus in cubic terms we may make the substitutions
 $$
 \stackrel{a}\tau{}^{\mu\nu}\to-2\stackrel{4}\tau{}^{\mu\nu};\quad
 \stackrel{b}\tau{}^{\mu\nu}\to-2\stackrel{3}\tau{}^{\mu\nu};\quad
 \stackrel{c}\tau{}^{\mu\nu}\to-\stackrel{3}\tau{}^{\mu\nu}-
 \stackrel{11}\tau{}^{\mu\nu};\quad
\stackrel{d}\tau{}^{\mu\nu}\to-\stackrel{2}\tau{}^{\mu\nu}-
 \stackrel{9}\tau{}^{\mu\nu};\quad
 $$     
   $$
 \stackrel{e}\tau{}^{\mu\nu}\to-\stackrel{1}\tau{}^{\mu\nu}-
 \stackrel{8}\tau{}^{\mu\nu};\quad
\stackrel{f}\tau{}^{\mu\nu}\to-\stackrel{3}\tau{}^{\mu\nu}-
 \stackrel{11}\tau{}^{\mu\nu};\quad
\stackrel{g}\tau{}^{\mu\nu}\to-\stackrel{2}\tau{}^{\mu\nu}-
 \stackrel{9}\tau{}^{\mu\nu};\quad
\stackrel{h}\tau{}^{\mu\nu}\to-\stackrel{1}\tau{}^{\mu\nu}-
 \stackrel{8}\tau{}^{\mu\nu};\quad
$$
$$
\stackrel{i}\tau{}^{\mu\nu}\to-\stackrel{5}\tau{}^{\mu\nu}-
 \stackrel{12}\tau{}^{\mu\nu};\quad
 \stackrel{j}\tau{}^{\mu\nu}\to-\stackrel{13}\tau{}^{\mu\nu}-
 \stackrel{15}\tau{}^{\mu\nu};\quad
\stackrel{k}\tau{}^{\mu\nu}\to-\stackrel{5}\tau{}^{\mu\nu}-
 \stackrel{12}\tau{}^{\mu\nu};\quad
\stackrel{l}\tau{}^{\mu\nu}\to-2\stackrel{9}\tau{}^{\mu\nu};\quad
$$
$$
\stackrel{m}\tau{}^{\mu\nu}\to-2\stackrel{12}\tau{}^{\mu\nu};\quad
\stackrel{n}\tau{}^{\mu\nu}\to-\stackrel{8}\tau{}^{\mu\nu}-
 \stackrel{10}\tau{}^{\mu\nu};\quad
\stackrel{o}\tau{}^{\mu\nu}\to-\stackrel{6}\tau{}^{\mu\nu}-
 \stackrel{13}\tau{}^{\mu\nu};\quad
 \stackrel{p}\tau{}^{\mu\nu}\to-2\stackrel{7}\tau{}^{\mu\nu};\quad
 $$
 $$
 \stackrel{q}\tau{}^{\mu\nu}\to-\stackrel{6}\tau{}^{\mu\nu}-
 \stackrel{13}\tau{}^{\mu\nu},\quad {\rm or}\quad
 \stackrel{q}\tau{}^{\mu\nu}\to-\stackrel{14}\tau{}^{\mu\nu}-
 \stackrel{15}\tau{}^{\mu\nu}.                                    \eqno(22)
 $$
 Noting that together with $\stackrel{h}\tau{}^{\mu\nu}
 \to-\stackrel{1}\tau{}^{\mu\nu}-
 \stackrel{8}\tau{}^{\mu\nu}$ we also have
 $\stackrel{h}\tau{}^{\mu\nu}\to-\stackrel{10}\tau{}^{\mu\nu}-
 \stackrel{16}\tau{}^{\mu\nu}$, we find
 $$
 \stackrel{16}\tau{}^{\mu\nu}\to\stackrel{1}\tau{}^{\mu\nu}+
 \stackrel{8}\tau{}^{\mu\nu}-\stackrel{10}\tau{}^{\mu\nu}.          \eqno(23)
 $$
( We note by the way that substitutions  $\tau\to{\cal T}$ in (22) and (23)
 give corresponding relations for cubic terms expressed in
 terms of $\bar h$).

 With the help of (22) and (23) we find that $t^{\mu\nu}$ in (17) can be
 substituted as follows
 $$
 t^{\mu\nu}\to\frac1{32\pi G}\sum_1^{16}e_s\stackrel{s}\tau{}^{\mu\nu}+
 \frac12(\stackrel{M}T{}^{\mu\alpha}h_{\alpha}{}^{\nu}+
 \stackrel{M}T{}^{\nu\alpha}h_{\alpha}{}^{\mu}),    \eqno(24)
$$
$e_s$ are given in Table 2.
 Now it easy to verify that the contribution to $c$ in (19)
 from the sum in (24) is zero:
 $$
 \sum_1^{16}e_sR_s=0.                                             \eqno(25)
 $$

 Using the linearized Einstein equation
 $$
 -h^{jn,\sigma}{}_{,\sigma}+h^{j\sigma}{}_{,\sigma}{}^{n}+
 h^{n\sigma}{}_{,\sigma}{}^{j}-h^{,jn}+\eta^{jn}(h_{,\sigma}{}^{\sigma}-
 h_{\sigma\lambda}{}^{,\sigma\lambda})=16\pi G\stackrel{M}T{}^{jn},   \eqno(26)
 $$
 we rewrite the interaction term in (24) in the form
 $$
 \frac12(\stackrel{M}T{}^{jn}h_n{}^k+\stackrel{M}T{}^{kn}h_n{}^j)=
 \frac1{16\pi G}[\stackrel{l}\tau{}^{jk}-\stackrel{m}\tau{}^{jk}-
 \stackrel{n}\tau{}^{jk}+\stackrel{o}\tau{}^{jk}
 -\stackrel{p}\tau{}^{jk}+\stackrel{q}\tau{}^{jk}].                  \eqno(27)
 $$
 For the Newtonian centers this interaction term is equal $2\phi
 \stackrel{M}T{}^{00}$. 

Now, using (22) we find
$$
h_{jk}\frac14(\stackrel{M}T{}^{jn}h_n{}^k+\stackrel{M}T{}^{kn}h_n{}^j)\to
\frac1{64\pi G}h_{jk}[-4\stackrel{6}\tau{}^{jk}+4\stackrel{7}\tau{}^{jk}+
2\stackrel{8}\tau{}^{jk}-4\stackrel{9}\tau{}^{jk}+
$$
$$
2\stackrel{10}\tau{}^{jk}+4\stackrel{12}\tau{}^{jk}-                  
4\stackrel{13}\tau{}^{jk}].                                       \eqno(28)
$$
These terms contribute 2 to $c$ in (19). Finally we have to take into 
account that $\stackrel{M}T{}^{jk}$ consist of the observable energy-momentum
and interaction part. We need here $\stackrel{M}T{}^{00}$. For it we have 
$$
\stackrel{M}T{}^{00}=\sum_am_a\frac{dx^0}{ds}\delta(\vec x-\vec
x_a(t))\approx \stackrel{M}T{}^{00 observ}(1+\frac12h_{00}).         \eqno(29)  
$$
This is because in the presence of gravitational field
$$
ds^2=-g_{00}dt^2(1-v^2).                                              \eqno(30)
 $$
Here $v^2$ is physical velocity, see \S 88 in [Landau and Lifshitz (1973)].
 Hence
 $$
\frac{dx^0}{ds}=\frac{1}{\sqrt{-g_{00}(1-v^2)}}\approx\frac{1}{\sqrt{1-v^2}}(1+
\frac12h_{00}).                                                      \eqno(31)
$$
 The interaction term $\frac12\stackrel{M}T{}^{00}h_{00}=
 -\phi\stackrel{M}T{}^{00}$ in (29) together with (27) makes $c=1$ in (19).
 Thus, we can obtain the desired value of $g_{00}$ using gravitational 
 energy-momentum tensor (11) in three graviton vertex (17).

 \section{Gravitational wave equation}

 If we want to consider the wave equation approach, we can't just add the
 gravitational energy-momentum tensor (11) to $\stackrel{M}T{}^{jk}$ in
 the wave equation (26). This would lead to an incorrect $g_{00}$.
 We must first "improve" $t^{jk}$ by adding some properly chosen expression,
 which do not affect the conservation laws. The simplest possibility is 
 to use $(\partial_j\partial_k-
 \eta_{jk}\partial_n\partial^n)\bar h_{\alpha\beta}\bar h^{\alpha\beta}$ or
 $(\partial_j\partial_k-
\eta_{jk}\partial_n\partial^n)\bar h^2$. The first possibility will lead to 
an  expression proportional to\\ $(\stackrel{5}{\cal T}{}^{jk}-
\stackrel{2}{\cal T}{}^{jk}-\stackrel{d}{\cal T}{}^{jk}+
\stackrel{k}{\cal T}{}^{jk}$). We discard this possibility because it 
would change the coefficient in front of $\stackrel{5}{\cal T}{}^{jk}$ 
in (11), which gives the correct energy-momentum tensor of 
gravitational plane wave, cf. \S 107 in [Landau and Lifshitz (1973)].

Using the second possibility, we add to $t^{jk}$ in (11) the expression
$$
\tilde t^{jk}=\frac1{32\pi G}(\stackrel{a}{\cal T}{}^{jk}-
\stackrel{f}{\cal T}{}^{jk}+\stackrel{4}{\cal T}{}^{jk}-
\stackrel{11}{\cal T}{}^{jk}).                                  \eqno(32)
$$
In terms of $\tau$ this expression has the same form, see (4)  and (5).
The constant factor on the r.h.s. in (32) is chosen so that $g_{00}$ 
has the desired form
(16). Thus, we obtain the "improved" gravitational energy-momentum tensor
$$
t_{jk}^{imp}=t_{jk}+\tilde t_{jk}.                               \eqno(33)
$$

Now, for simplicity we consider a single Newtonian center. Outside the matter
we find
$$
 \tilde t_{00}=-\frac{GM^2}{2\pi r^4},\quad \tilde t_{jk}=
 \frac{GM^2}{\pi}\left(\frac{\delta_{jk}}{r^4}-\frac{2x_jx_k}{r^6}\right),
                                                    \eqno(34)
 $$
 and for $t_{\mu\nu}$ in (11)
 $$
 t_{00}=\frac{GM^2}{8\pi r^4},\quad  t_{jk}=
\frac{GM^2}{8\pi}\left(-\frac{\delta_{jk}}{r^4}+\frac{2x_jx_k}{r^6}\right).
                                                     \eqno(35)
 $$
Beginning from equation (34) in all equations up to equation (46) 
$j,k=1,2,3$. From (32)-(35) it follows 
 $$
 t_{00}^{imp}=-\frac{3GM^2}{8\pi r^4},\quad
 t_{jk}^{imp}=
 \frac{7GM^2}{8\pi}\left(\frac{\delta_{jk}}{r^4}-\frac{2x_jx_k}{r^6}\right).
                                                   \eqno(36)
$$

We note that $t_{jk}$, $\tilde t_{jk}$, and $t_{jk}^{imp}$, 
differ only by constant factors and
$$
\left(\frac{\delta_{jk}}{r^4}-\frac{2x_jx_k}{r^6}\right)_{,k}=0.    \eqno(37)
$$
Thus, the conservation laws $t^{\mu\nu}_{\nu}=0$ are satisfied.

Comparison of $t^{imp}_{\mu\nu}$ in (36) with corresponding Weinberg tensor
 ( see equations (7.6.3) and (7.6.4) in [Weinberg (1972)])
shows that for the considered special case they coincide: $t_{\mu\nu}^{imp}=
t_{\mu\nu}^W$, if (7) is used for calculating $t_{\mu\nu}^W$.
 The expressions for $g_{jk}$ in Hilbert 
($\bar h^{jk}_{,\nu}=0$), harmonic and isotropic frames are 
$$
g_{jk}^{Hilb}=\delta_{jk}(1-2\phi)+\phi^2(5\delta_{jk}-\frac{7x_jx_k}{r^2}),
\quad g_{jk}^{har}=\delta_{jk}(1-2\phi)+\phi^2(\delta_{jk}+\frac{x_jx_k}{r^2}),
$$
$$
 g_{jk}^{iso}=\delta_{jk}(1-2\phi+\frac32\phi^2). \eqno(38)
$$
In all these frames $g_{00}$ are the same as in (16). The differences in
$g_{jk}$ are proportional to
$$
\Lambda_{i,k}+\Lambda_{k,i}=
2(\left(\frac{\delta_{jk}}{r^2}-\frac{2x_jx_k}{r^4}\right))\quad
     \Lambda_i=\frac{x_i}{r^2},                                 \eqno(39)
$$
i.e. these $g_{jk}$ are related by gauge transformations. This is in agreement 
with the fact that in considered approximation Weinberg gravitational energy-momentum
tensors in harmonic and isotropic frames coincide [Nikishov (2003)].

Defining as usual $\bar t_{\mu\nu}=t_{\mu\nu}-\frac12\eta_{\mu\nu}t$, we
get from (36)
$$
\bar t_{00}^{imp}=\frac{GM^2}{4\pi r^4};\quad \bar t_{jk}^{imp}=
  \frac{GM^2}{4\pi}\left(\frac{\delta_{jk}}{r^4}-\frac{7x_jx_k}{r^6}\right).
                                                   \eqno(40)
  $$

Using improvement technique we can make any gravitational energy-momentum 
tensor suitable for using in the wave equation. In this way we obtain
another version of theory in considered approximation. For example,
for Thirring (1961) tensor we have instead of (35) 
$$
t_{00}^{Th}=-\frac{7GM^2}{8\pi r^4};\quad t_{jk}^{Th}=\frac{GM^2}{8\pi}
\left( \frac{2x_jx_k}{r^6}-\frac{\delta_{jk}}{r^4}\right).           \eqno(41)
$$
The "improved" tensor has the form
$$
t_{\mu\nu}^{imp Th}=t_{\mu\nu}^{Th}+3\tilde t_{\mu\nu},   \eqno(42)
$$
where $\tilde t_{\mu\nu}$ is given in (34).
From (41), (42) and (34) we find
$$
 t_{00}^{imp Th}=-\frac{19GM^2}{8\pi r^4};\quad t_{jk}^{imp Th}=
 \frac{23GM^2}{8\pi}\left(-\frac{2x_jx_k}{r^6}+\frac{\delta_{jk}}{r^4}\right).
                                                                   \eqno(43)
 $$
 From here for barred quantities we get
 $$
 \bar t_{00}^{imp Th}=\frac{GM^2}{4\pi r^4};\quad\bar t_{jk}^{imp Th}=
 \frac{GM^2}{4\pi}\left(-\frac{23x_jx_k}{r^6}+\frac{\delta_{jk}}{r^4}\right).
                                                                   \eqno(44)
 $$
 As seen from (40) and (44) $\bar t_{00}^{imp Th}=\bar t_{00}^{imp}$.
 Hence, $g_{00}$ are the same in both versions, but $g_{jk}$ are
 different. Using the wave equation in the form
 $$
 -h_{\mu\nu,\sigma}{}^{\sigma}+\bar h_{\mu\sigma}{}^{\sigma}{}_{\nu}+
 \bar h_{\nu\sigma}{}^{\sigma}{}_{\mu}=16\pi G\bar T_{\mu\nu},     \eqno(45)
 $$
we find  in Hilbert gauge ($\bar h^{\mu\nu}{}_{,\nu}=0$) 
   $$
 g_{jk}^{Th}=\delta_{jk}(1-2\phi)+G^2M^2
 \left(-\frac{23x_jx_k}{r^6}+\frac{21\delta_{jk}}{r^4}\right).     \eqno(46)
 $$

 In conclusion of this Section we note that the nonuniqueness of gravitational
 energy-momentum tensor in general relativity is connected with the
 nonphysical degrees of freedom taking part in its formation [Nikishov (2003)].
  If we could
 define the preferred frame as the one formed by only physical
 degrees of freedom, then the problem of uniqueness could be solved.
 It seems reasonable to assume that for the Newtonian center in linear
 approximation the preferred metric is given by (7) and in the second
 approximation it may be the harmonic one.

 \section{Remark on Feynman gravitational Lagrangian}

 The Feynman approach to building up gravity theory seems to me quite 
 natural because he uses throughout $h_{\mu\nu}$ as gravitational variable,
 not some function of it. His cubic term of the Lagrangian 
 (see equation (6.1.13) in [Feynman, Moringo, Wagner (1995)]) in our
  notation  has the form
 $$
 F^3=-\frac1{64\pi G}h^{jk}\{\frac18\stackrel{a}\tau{}_{jk}+
 \frac14\stackrel{b}\tau{}_{jk}-\frac12\stackrel{c}\tau{}_{jk}-
 \frac12\stackrel{g}\tau{}_{jk}+\stackrel{i}\tau{}_{jk}-
 \frac34\stackrel{l}\tau{}_{jk}+\frac12\stackrel{m}\tau{}_{jk}+
 \stackrel{n}\tau{}_{jk}-2\stackrel{o}\tau{}_{jk}+
 \stackrel{p}\tau{}_{jk}
 $$
 $$
 +(\stackrel{3}\tau{}_{jk}-\frac14\stackrel{4}\tau{}_{jk}-
 2\stackrel{10}\tau{}_{jk}+\frac12\stackrel{11}\tau{}_{jk}+
 2\stackrel{14}\tau{}_{jk}-\stackrel{16}\tau{}_{jk})\}.              \eqno(47)
 $$
 This expression have to be compared with cubic term in the expansion of 
 general relativity Lagrangian $-\frac1{16\pi G}\sqrt{-g}R$.
 My calculation gives the following result for this term
 $$
 -\frac1{16\pi G}h^{jk}\{\frac18\stackrel{a}\tau{}_{jk}-
 \frac18\stackrel{b}\tau{}_{jk}-\frac14\stackrel{c}\tau{}_{jk}-
 \frac14\stackrel{f}\tau{}_{jk}-
 \frac12\stackrel{g}\tau{}_{jk}+\stackrel{h}\tau{}_{jk}-
 \stackrel{i}\tau{}_{jk}-\stackrel{j}\tau{}_{jk}-
 \frac14\stackrel{l}\tau{}_{jk}+\frac14\stackrel{m}\tau{}_{jk}+
 \stackrel{n}\tau{}_{jk}
 $$
 $$
 -\frac12\stackrel{o}\tau{}_{jk}+
 \stackrel{p}\tau{}_{jk}-\frac32\stackrel{q}\tau{}_{jk}+
 (\frac14\stackrel{1}\tau{}_{jk}-\frac38\stackrel{2}\tau{}_{jk}-
 \frac12\stackrel{3}\tau{}_{jk}+\frac18\stackrel{4}\tau{}_{jk}+
 \frac34\stackrel{5}\tau{}_{jk}-\frac12\stackrel{6}\tau{}_{jk}+
 $$
 $$
 \frac32\stackrel{7}\tau{}_{jk}+\stackrel{8}\tau{}_{jk}-
 \frac12\stackrel{9}\tau{}_{jk}+\stackrel{10}\tau{}_{jk}-
 \frac14\stackrel{11}\tau{}_{jk}+\stackrel{12}\tau{}_{jk}-
 \stackrel{13}\tau{}_{jk}-\stackrel{14}\tau{}_{jk}-
 2\stackrel{15}\tau{}_{jk}+\frac12\stackrel{16}\tau{}_{jk})\}   \eqno(48)
 $$
 Using reductions to first derivatives (22) and equation (23), 
 we get for (48) the 
 expression (18). For Feynman's Lagrangian (47) the reduction gives
 (18) with substitution $4\stackrel{13}\tau{}_{jk}\to2\stackrel{13}\tau{}_{jk}+
 2\stackrel{14}\tau{}_{jk}$.Thus, it seems that Feynman's Lagrangian 
 (47) differs essentially from (48). Evidently Feynman do not think so. In any
 case his Lagrangian, used as a vertex,  explain perihelion precession. This is seen from the fact 
 that for Newtonian centers $\stackrel{13}\tau{}_{jk}=
 \stackrel{14}\tau{}_{jk}$, see (9) and text below Table 2.

 \section{Conclusion}

 Our considerations show that the ways are open in search for
  gravitation theory with unique gravitational energy-momentum tensor,
  giving the positive energy density. It is still not clear
  which coordinate condition better
  exclude the nonphysical degrees of freedom; is it harmonic, Hilbert or some
  other condition? Will the future gravitational theory be the theory of
   gravitons (spin-2), or gravitons and spin-0 field (as in 
   [Baryshev (1999)] and [Logunov (1998)])?

 \section {Acknowledgements}
 I am grateful to V.I.Ritus, R.Metsaev and A.Barvinsky for
  valuable discussions and suggestions.   
 This work was supported in part by the Russian Foundation for
Basic Research (projects no. 00-15-96566 and 02-02-16944).
  
   \section*{References} 
  Baryshev Yu., gr-qc/9912003.   \\
  Cohn J.,Int. Jour. of Theor. Phys. {\bf2}, 267 (1969).\\  
  Dehnen H., H\"onl H., and Westpfahl K., Ann.  der
          Phys. {\bf6}, 7 Folge, Band 6, Heft 7-8, S.670 (1960).\\
  Einstein A., Ann. Phys., {\bf49}, 769 (1916).\\ 
  Feynman R.P., Moringo F.B., Wagner W.G.,
        {\sl Feynman Lectures on Gravitation}, Addison-Wesley Company (1995).\\
  Iwasaki Y., Prog. Theor. Phys.  {\bf46}, 1587 (1971).  \\
   Landau L.D. and Lifshitz E.M., {\sl The classical theory of
                         fields}, Moscow, (1973) (in Russian).\\
  Logunov A.A., Part. and Nucl. {\bf29} (1) Jan.-Feb 1998, p 1;
  {\sl The Theory of Gravitational Field}, Moscow, Nauka (2000), (in Russian).\\
   Misner C.W., Thorne K.S., Wheeler J.A., {\sl Gravitation.} San
                        Francisco(1973).\\
  Nikishov A., gr-qc/9912034; Part. and Nucl. {\bf32}, 5 (2001).\\
  Nikishov A., gr-qc/0310072\\ 
  Schwinger J. {\sl Particles, Sources, and Fields.} V.1 Addison-Wesley
                             (1970).\\
  Thirring W.E., Ann. Phys. (N.Y.) {\bf16}, 96 (1961).  \\
   Wald R.M., Phys. Rev. {\bf33,} 3613,(1986).       \\
  Weinberg S.,{\sl Gravitation and Cosmology}, New York (1972).\\
\end{document}